\documentclass[aps,pre,twocolumn,floatfix,superscriptaddress]{revtex4}

\usepackage{graphicx}  
\usepackage{dcolumn}   
\usepackage{bm}        
\usepackage{amssymb}   
\usepackage[latin1]{inputenc}
\usepackage{amsfonts}
\usepackage{amsmath}
\usepackage{amsthm}
\usepackage{xcolor}
\usepackage{fontenc}
\usepackage{textcomp}
\usepackage{epstopdf}
\usepackage{hyperref}

\begin{document}
\title{Effect of shape anisotropy on percolation of aligned and overlapping objects on lattices}
\author{Jasna C. K.}
\email{jasnack@cusat.ac.in}
\affiliation{Department of Physics, Cochin University of Science and Technology, Cochin 682022, India.}
\author{V. Sasidevan}
\email{sasidevan@cusat.ac.in, sasidevan@gmail.com}
\affiliation{Department of Physics, Cochin University of Science and Technology, Cochin 682022, India.}
\date{\today}

\begin{abstract}
{\small We investigate the percolation transition of aligned, overlapping, anisotropic shapes on lattices. Using the recently proposed lattice version of excluded volume theory, we show that shape-anisotropy leads to some intriguing consequences regarding the percolation behavior of anisotropic shapes. We consider a prototypical anisotropic shape - rectangle - on a square lattice and show that for rectangles of width unity (sticks), the percolation threshold is a monotonically decreasing function of the stick length, whereas, for rectangles of width greater than two, it is a monotonically increasing function. Interestingly, for rectangles of width two, the percolation threshold is independent of its length. We show that this independence of threshold on the length of a side holds for $d-$dimensional hypercubiods as well for specific integer values for the lengths of the remaining sides. The limiting case of the length of the rectangles going to infinity shows that the limiting threshold value is finite and depends upon the width of the rectangle. 
This \textquoteleft continuum\textquoteright\; limit with the lattice spacing tending to zero only along a subset of the possible directions in $d-$dimensions results in a novel \textquoteleft semi-continuum\textquoteright\; percolation system.
We show that similar results hold for other anisotropic shapes and lattices in different dimensions. The critical properties of the aligned and overlapping rectangles are evaluated using Monte Carlo simulations. We find that the threshold values given by the lattice-excluded volume theory are in good agreement with the simulation results, especially for larger rectangles. We verify the isotropy of the percolation threshold and also compare our results with models where rectangles of mixed orientation are allowed. Our simulation results show that alignment increases the percolation threshold. The calculation of critical exponents places the model in the standard percolation universality class. Our results show that shape-anisotropy of the aligned, overlapping percolating units has a marked influence on the percolation properties, especially when a subset of the dimensions of the percolation units are made to diverge.} 
\end{abstract}
\maketitle
\section{INTRODUCTION}
Percolation problems have commanded enduring interest in the realm of statistical physics, finding diverse applications in fields ranging from material science and polymer chemistry to epidemic spreading and financial markets \cite{sahimi,sander,jiantong,li,drossel,klypin,adam,stauffer1}. The fundamental focus of percolation theory lies in investigating the connectivity properties of random and disordered media. In the simplest model of lattice site percolation, each site of a lattice is randomly occupied with a probability $p$. Clusters are formed by the nearest occupied sites, and as the occupation probability is gradually increased from zero, the system eventually reaches a point where the largest cluster spans the entire lattice, signaling the occurrence of percolation \cite{christensen, stauffer}. Percolation models can also be extended to continuum space, where geometric shapes or objects, which can partially or fully overlap, are randomly placed in the space with a specific number density. Some of the commonly considered shapes include discs, spheres, cubes, squares, and sticks \cite{mertens, xia, baker}. Overlapping objects give rise to distinct clusters, and, similar to lattice percolation, the system exhibits a phase transition, marked by the emergence of a spanning cluster at a critical number density. Identifying the percolation threshold and characterizing the associated critical behavior constitute a major research theme in the study of various percolation systems \cite{stauffer}. The probability of forming a spanning cluster serves as one of the order parameters, and the critical phenomenon is characterized by a power-law divergence of specific quantities close to the percolation threshold. Also, the universality of critical exponents is seen across many models of percolation \cite{gawlinski}.

Among the diverse variants of percolation on lattices, many recent studies have focused on the percolation of extended shapes that do not overlap \cite{becklehimer,vandewalle,longone,tarasevich,tarasevich1,cornette,cornette1,lebrecht,grzegorz,grzegorz1,cherkasova,longone1}. Compared to this, there are only a few studies on extended shapes that overlap \cite{koza,koza1,brzeski,mecke}. The models with overlapping shapes introduce multisite occupancy and/or multiple occupancy of a site, also serving as a natural bridge between various lattice and continuum models. For example, the percolation of overlapping squares and cubes on lattices was studied by Koza et al. \cite{koza}, who focused on calculating the percolation thresholds of these models and their transition from discrete to continuum values. Percolation of discrete overlapping hyperspheres on hypercubic lattices was studied by Brzeski and Kondrat \cite{brzeski}. They studied the discrete-to-continuum transition of hyperspheres, which enabled the evaluation of the threshold for the 3D and higher dimensional continuum problems with greater accuracy. Apart from these, recent studies focus on site percolation with extended neighborhoods, which can be mapped onto the problem of lattice percolation of extended and overlapping shapes like discs, squares, etc \cite{koza, xun, majewski}.

Percolation models involving overlapping shapes, in addition to their theoretical significance, hold practical relevance in various applications. These models, applied to both lattice and continuum structures, serve as valuable tools for representing the structure of random and disordered media and investigating their diverse transport properties. The collective arrangement of overlapping shapes can represent either the material components or voids within a structure. For example, overlapping ellipses have been used for modeling composites of conductive nanoparticles dispersed in insulating matrices \cite{alvarez}. Similarly, electrical conductivity in nano-composites has been studied using both soft-core (non-interacting) and hard-core (interacting) shapes \cite{balberg2}. Other examples include the study of nano-composite materials consisting of conductive rod-like particles \cite{du, ni}, systems consisting of conducting polymers \cite{ramasubramaniam,hu}, and composite fiber systems \cite{berhan,berhan1}. The latter gives a comparison between soft-core and hardcore models for fiber systems. Numerous real-world scenarios involve the flow of fluids or other conductive phenomena, such as electrical conductivity, in 2D structures \cite{adler,saleh,yaofa,tang}. In the past, studies have utilized 2D overlapping aligned rectangles, integrated with an underlying lattice structure, to analyze transport properties \cite{koponen,koponen1,koponen2,matyka,koza2}. Apart from these examples, the fact that image analysis of a disordered material usually involves discretizing a continuous space, makes the studies on lattices very much relevant in the context of random media \cite{mecke1}. Thus it is evident that the versatility and applicability of percolation models with overlapping shapes, both in 2D and 3D lattice structures make them pertinent in understanding and characterizing various physical systems.

This paper delves into the percolation of overlapping and aligned anisotropic shapes on lattices. Specifically, as a prototypical example of such problems, we consider the percolation of aligned and overlapping \textquotedblleft rectangles\textquotedblright\; on a two-dimensional square lattice in detail. In the particular case of either side of the rectangle being of unit length, we can call the shapes rods or sticks (See Fig.~\ref{figure1}). Apart from their potential practical relevance, our main motivation behind considering this and other extended anisotropic shapes stems from the peculiar behavior of the percolation threshold of aligned and overlapping rectangles in the 2D continuum problem. In the latter, the percolation threshold remains independent of the rectangles' aspect ratio due to the affine symmetry of such systems \cite{klatt}. In simple words, for such systems, scaling one or both directions will not change the connectivity properties of the system and, hence, will not change the percolation threshold. Similar invariance of percolation threshold under scaling also exists for disc percolation in continuum \cite{dhar}. However, on lattices, such symmetry under scaling cannot be defined, and we show that this leads to intriguing percolation behavior, which depends not just on the aspect ratio but on the specific values of the side-lengths of the aligned rectangles and other shapes.

Employing the excluded volume theory adapted to a lattice setting \cite{koza} and employing Monte Carlo simulations, we analyze the percolation of aligned and overlapping rectangles on the two-dimensional square lattice. We show that the lattice version of the excluded volume theory gives several intriguing predictions regarding the percolation threshold of such systems. For the specific case of aligned and overlapping rectangles, the value of the percolation threshold as we increase the length of the rectangles depends on its width. For width one rectangles, the threshold monotonically decreases with the length, and for width greater than two, it increases monotonically. For width two, the threshold is independent of the length!.  This independence of threshold on the length of a side holds for $d-$dimensional hypercuboids as well for specific integer values for the lengths of the remaining sides. Moreover, we get a non-zero percolation threshold, even in the limiting case where one side of the rectangle extends to infinity while keeping the other side finite. We show that similar results hold for other shapes and dimensions as well and obtain predictions for both qualitative and quantitative behavior of the threshold for several such systems. The theory gives us good numerical estimates for the percolation thresholds for rectangles and other anisotropic shapes.

Our Monte Carlo simulation results confirm the theoretical predictions regarding the behavior of the percolation thresholds for aligned and overlapping rectangles. We verify that isotropy of the percolation threshold holds even with rectangles of large aspect ratios, just as in the continuum problem \cite{klatt}. Additionally, critical exponents are obtained, demonstrating that the problem lies within the same universality class of lattice percolation. Furthermore, a comparison between results for aligned sticks and a mixture of sticks of varying orientations is presented, providing valuable insights into the impact of anisotropy on the percolation process. 

The paper is organized as follows. In section \ref{sec1}, we precisely define the model of overlapping rectangles on a square lattice. In Section \ref{sec2}, we use excluded volume theory adapted to a lattice setting to analyze the problem. We use the theory to draw conclusions about the percolation of extended shapes other than rectangles and also in other dimensions. We give the results of simulation studies. The percolation threshold is evaluated, and results are compared. Isotropy of percolation threshold, critical exponents, and effect of mixed orientations are also given. We conclude in section \ref{sec3}.

\section{MODEL DEFINITION}
\label{sec1}
\begin{figure*}[hbtp]
\includegraphics[scale=1]{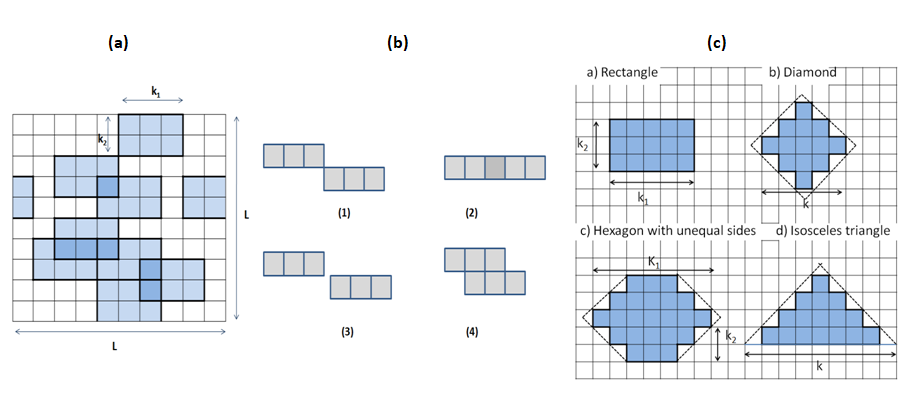}
\caption{ (a) Square lattice of size $L\times L$ with overlapping rectangles of length $k_1 = 3$ and width $k_2 = 2$ randomly distributed on it. Rectangles are aligned in the horizontal direction and can overlap. (b) Different possible arrangements of two rectangles with $k_1=3$ and $k_2=1$ (1) Two rectangles touching at corners are considered unconnected. (2) Two overlapping rectangles are considered connected (3) Two non-overlapping rectangles (4) Two rectangles sharing edges are considered connected. (c) A few of the extended shapes possible on a square lattice are shown with their defining parameters marked. The name for a shape is given based on the shape of its outline.}
\label{figure1}
\end{figure*}

In general, we look into the percolation problem of aligned, overlapping, anisotropic shapes in lattice systems. For concreteness, we consider the case in which each unit of percolating objects is a rectangle with side lengths $k_1$ in the horizontal direction and $k_2$ in the vertical direction randomly positioned on a two-dimensional square lattice of size $L \times L$ (See Fig. \ref{figure1}). Without loss of generality, we will assume that $k_1 \geq k_2$ so that the rectangles are aligned in the horizontal direction. 

The rectangles are allowed to overlap; hence, a site can be occupied more than once - a feature known as multiple occupancy. However, the complete overlapping of two rectangles is avoided (Note that the second rectangle doesn't contribute toward the total number of occupied sites). This will also retain the classical site percolation scenario when $k_1 = k_2 = 1$. Occupied neighboring sites are assumed to be connected, where the neighborhood is the Von-Neumann type. In other words, rectangles that share sites or are adjacent are considered connected entities, and rectangles that touch only at corners are considered non-connected (See Fig. \ref{figure1}(a) \& \ref{figure1}(b)). 

\begin{figure*}[hbtp]
\centering
\includegraphics[scale=0.75]{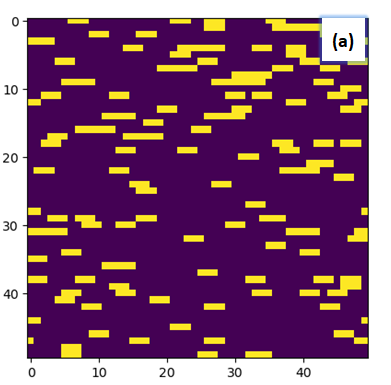}
\includegraphics[scale=0.75]{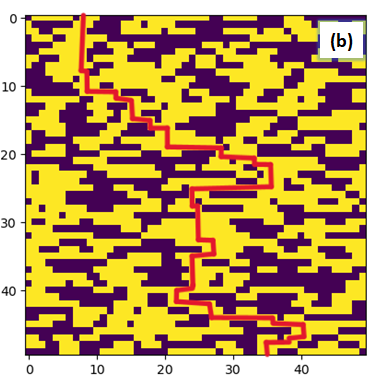}
\includegraphics[scale=0.75]{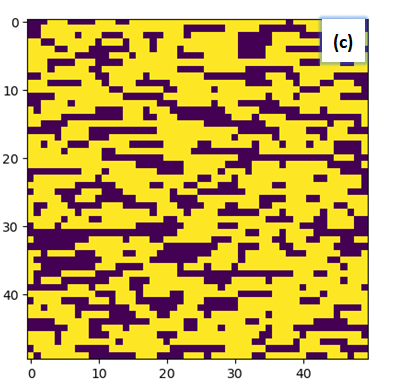}
\caption{Typical configurations obtained for rectangles of side lengths $k_1=3, k_2 = 1$ distributed randomly on a $L= 50 \times 50$ square lattice for different values of the density of occupied sites $\phi$. The lighter color corresponds to the rectangles. The system is considered as percolating if there is connectivity from top to bottom via rectangles. (a) At $\phi = 0.14 < \phi_c$, system is non-percolating. (b) At $\phi = 0.56 \approx \phi_c$, the system is percolating with the first spanning cluster developed in the vertical direction. A percolating path connecting the top to the bottom is marked. (c) At $\phi = 0.64 $, the system is well above the percolation threshold $\phi_c$.}
\label{figure2}
\end{figure*}

As the density of occupied sites, denoted by $\phi$ (the ratio of the total number of occupied sites and $L^2$), is progressively increased by adding more rectangles to the lattice, a critical point is reached where a spanning cluster emerges. We define the spanning cluster as a connected path of rectangles in the vertical direction (perpendicular to the direction of alignment), signifying percolation in the system. Periodic boundary conditions are imposed in the horizontal direction. Snapshots of the typical configurations for different values of $\phi$, with $k_1 = 3$ and $k_2 = 1$, are shown in Fig.~\ref{figure2}. 

The percolation threshold is the specific value of $\phi$ where the spanning cluster first arises, indicating the transition from a non-percolating to a percolating state. $\phi_c$ may be considered the equivalent of the critical covered volume fraction (CCVF) in continuum percolation \cite{mertens}. In continuum percolation theory, we have areal density $\eta$ and covered area/volume fraction related to each other through $\phi=1-\exp\left(-\eta\right)$ \cite{mertens}. Areal density is the net volume of all the objects per unit volume of the space, and covered area fraction is the fraction of the total volume of space covered by the objects. The objects are randomly distributed in space; hence, the probability that there are a certain number of objects in an area follows a Poisson distribution. Hence, the probability that a point in space is not covered by any object is the same as the probability that there are no objects present within a volume equal to the volume of the object $V$ (or average volume of objects if there is a size distribution for the objects) which is $\exp\left(-V n\right)$ where $n$ is the number density of objects. $\eta  = V n$ is the areal density. 
$\phi$ is then the probability that the point in space is covered by at least one object, which is $1 - \exp(-\eta)$. 
It is easily seen that the same arguments and relation hold for extended overlapping objects on lattices as well, with $n$ now denoting the number density of objects on the lattice and $V$ the 'volume' of the object (the number of lattice sites occupied by one object). $\eta$ is the equivalent of the area/volume density in continuum percolation. For convenience, we will use this terminology for the lattice percolation as well. 

Note that when $k_1 = k_2 = k \rightarrow \infty$, the problem corresponds to the 2D continuum percolation problem of overlapping and aligned squares \cite{koza, koza1}.  The problem may be defined with shapes other than rectangles and also for other types of lattices as well. For example, a few simple shapes that can be considered on the square lattice are shown in  Fig.~\ref{figure1}(c). We will discuss other extended shapes, lattices, and dimensions later in Sec.\ref{sec2}. 

\section{RESULTS AND DISCUSSION}
\label{sec2}
\subsection{Lattice version of excluded volume theory}
\label{subsec1}
A major quantity of interest in the problem is the percolation threshold. i.e, the value of $\phi$ at which the system percolates for given values of $k_1$ and $k_2$.
A useful analytical approximation technique to obtain the percolation threshold of systems for which the shapes of the percolating units are similar is the excluded volume theory \cite{balberg, balberg1}. The excluded volume theory, initially proposed for continuum percolation systems, is based on the idea that the product of the number density of basic percolating units and the average excluded volume, which gives us the total excluded volume, is an invariant quantity for \textquoteleft similar systems\textquoteright\; at the critical point. Similar systems here mean geometrically similar shapes that are distributed with a particular orientation \cite{balberg}. For example, systems of squares and rectangles of a particular orientation. In the continuum context, this means shapes that can be scaled to each other and have the same orientation. For shapes considered on a lattice, since there is no notion of scaling, we can say that two shapes are similar if their continuum limit can be scaled to each other and have the same orientation. Therefore, we expect the total excluded volume/area to be the same for systems of oriented squares and oriented rectangles (See Ref.~\cite{xun1} for similar findings in a related context).

\begin{figure}
   \includegraphics[scale=0.6]{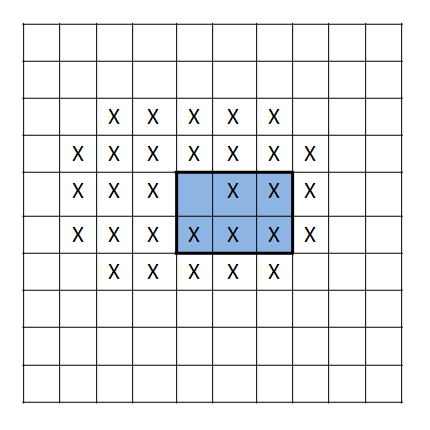}
    \caption{If the top-left corner site of the rectangle is chosen as the index site (Site with no cross mark), the cross-marked cells constitute the excluded area of the rectangle shown, which is of length $k_1=3$ and width $k_2=2$. i.e., if the index site of another rectangle falls within this area, it will overlap with the rectangle shown in the figure. The excluded volume is $V_{ex}= \left(2k_1+1\right)\left(2k_2+1\right) - 5$. Note that exact overlapping of rectangles is not allowed.}
    \label{Vex}
\end{figure}

For continuum cases, the excluded volume is the volume around an object into which, if another object is placed, the two objects will overlap. Recently, the idea has been extended to lattice models as well, where the excluded volume is replaced by what is called the connectedness factor - the number of possible configurations of two basic percolating units such that they are connected - \cite{koza} (See Fig.~\ref{Vex}). The connectedness factor will depend on the shape of the objects and also on their relative orientation.  The lattice version of excluded volume theory thus says that the product of the number density of objects and the connectedness factor is a constant at criticality for similar systems where similarity refers to the shape and orientation of the basic percolating units. If $n$ is the number density of objects (total number of objects divided by the total number of lattice points) and $V_{ex}$ is the connectedness factor, then 
\begin{equation}
n_{c}V_{ex}=B_{c}
\label{eq1}
\end{equation}
where $n_c$ is the number density at the critical point and $B_{c}$ is the average number of connections an object has at criticality, which is approximately expected to take the same value for similar objects.
Here, the number density $n=\eta/V$ where $V$ is the volume (number of lattice sites occupied) of an object. 
Therefore, at criticality, we can write the CCVF,
\begin{equation}
\phi_{c} \approx 1-\exp \left(-B_{c}\frac{V}{V_{ex}}\right)
\label{eq2}
\end{equation}
Now for rectangles of sides $k_1$ and $k_2$, $V$ is simply $k_1 \times k_2$. We can find $V_{ex}$ by enumerating the configuration of two rectangles such that they are connected. It is easily seen that  $ V_{ex} = \left(2k_1+1\right)\left(2k_2+1\right) - 5$ (See Fig.~\ref{Vex}). Therefore, the covered volume fraction at the critical point,
\begin{equation}
\phi_{c}^{k1,k2} \approx 1-\exp\left(-B_{c}\frac{k_{1}k_{2}}{(2k_{1}+1)(2k_{2}+1)-5}\right)
\label{eq3}
\end{equation}

So, unlike the continuum case, here the CCVF depends on the aspect ratio of the rectangles. We can recover the continuum case in the limit of both $k_1$ and $k_2$ much greater than unity for which the above expression becomes independent of $k_1$ and $k_2$,
\begin{equation}
\phi_{c} \approx 1-\exp(-B_c/4) 
\label{eq4}
\end{equation}

Several interesting inferences can be made from Equation~(\ref{eq3}). Consider the case in which we take $k_1$ to be very large for a fixed finite value of $k_2$. In the limit of $k_1 \rightarrow \infty$, the expression for CCVF becomes,
\begin{equation}
\phi_{c}^{k_1 \rightarrow \infty, k_2} \approx 1-\exp\left(-B_{c}\frac{k_{2}}{2(2k_{2}+1)}\right)
\label{eq5}
\end{equation}
Equation~(\ref{eq3}) gives how the percolation threshold approaches the limiting value in Equation.~(\ref{eq5}) as we increase $k_1$. We can easily see that for $k_2 = 1$, $\phi_c^{k_1,k_2}$ is a decreasing function of $k_1$ whereas for $k_2 > 2$, $\phi_c^{k_1,k_2}$ is increasing in $k_1$. Now for $k_2 = 2$, $\phi_c^{k_1,k_2}$ is independent of $k_1$, indicating that for overlapping and aligned rectangles of width two on a square lattice, the percolation threshold is independent of its length!

Equation (\ref{eq5}) implies that for rectangles with finite width and length tending to infinity, the percolation threshold is finite and depends on the width of the rectangles. Moreover, unlike squares whose infinite limit gives us the continuum percolation problem of squares, there seems to be no continuum problem corresponding to the case of rectangles with infinite length and finite width. This is true as long as we consider the case $k_1\rightarrow \infty$ keeping $k_2$ a constant. We will discuss the transition to the continuum picture in more detail in section \ref{subsubsec4}.

It is interesting to note that for the particular case of aligned overlapping sticks ($k_2 = 1$), the dependence of percolation threshold on stick length $k_1$ is given by $\phi_{c}~\sim 1~-~\exp\left(-B_c \dfrac{k_1}{3\left(2k_1+1\right)~-~5}\right)$. We may compare and contrast this with the case of non-overlapping stick percolation. The latter system is often predicted to have a power-law dependence of the percolation threshold on stick length $\phi_{c} \sim 1/k_1^{\alpha}$ \cite{becklehimer, tarasevich1} or an exponentially decreasing dependence \cite{cornette1}. 

We can easily see that similar results will hold for extended shapes in other types of lattices and dimensions. For example, consider the problem of overlapping and aligned cuboids on a cubic lattice. If the side lengths of the cuboids are denoted by $k_1$, $k_2$ and $k_3$, then CCVF is given by,

\begin{equation}
\phi_{c}^{k1,k2,k3} \approx 1-\textrm{exp}\left(-B_c\frac{k_{1}k_{2}k_{3}}{V_{ex}^{k_1,k_2,k_3}}\right)
\label{eq6}
\end{equation}

where $V_{ex}^{k_1,k_2,k_3}$ is,
\begin{align}
V_{ex}^{k_1,k_2,k_3} =~&\left(2k_{1}-1\right)\left(2k_{2}-1\right)\left(2k_{3}-1\right)  + \notag \\ &2[\left(2k_{1}-1\right)\left(2k_{2}-1\right)+\left(2k_{1}-1\right)\left(2k_{3}-1\right)+ \notag \\&\left(2k_{2}-1\right)\left(2k_{3}-1\right)]-1
\label{eq7}
\end{align}

When all the side lengths tend to infinity, we recover the continuum percolation threshold of aligned and overlapping cubes in $3D$, given by 
\begin{equation}
    \phi_c = 1 - \exp\left(\dfrac{-B_c}{8}\right)
\label{eq8}
\end{equation}
 with appropriate values of $B_c$. From Equation~(\ref{eq7}) and Equation~(\ref{eq6}), we can verify that for $k_2=2$ and $k_3=4$ (or vice versa), the percolation threshold $\phi_c$ becomes independent of $k_1$. We can easily show that this set of values is unique. In Equation~(\ref{eq7}), if we set $a_2= (2k_2-1)$ and $a_3= (2k_3-1)$, for $\phi_c$ to be independent of $k_1$, we require to have odd positive integer values of $a_2$ and $a_3$ greater than or equal to 3 that satisfy the relation $a_2=\frac{2a_3+1}{a_3-2}$ (or $a_3=\frac{2a_2+1}{a_2-2}$). Now if $a_3 = 3$, then $a_2 = 7$. For higher values of $a_3$, the ratio $\frac{2a_3+1}{a_3-2}$ is monotonically decreasing and tends to 2 for large $a_3$. So for a higher value of $a_3$ to satisfy the above relation, the only remaining possibility is $a_2 = 5$. However, it is easy to verify that $a_2 = 5$ will not satisfy the relation for any integer value of $a_3$. This shows that the combinations $a_2=3$ and $a_3=7$ ($k_2 = 2, k_3 = 4$) or $a_2=7$ and $a_3=3$ ($k_2 = 4, k_3 = 2$) are the only possible set of integers which will make the threshold independent of $k_1$. We can generalize this to hyper-cuboids in $d-$dimensions where we expect $\phi_c$ to be independent of $k_1$ for a particular set of integer values of $k_2$, $k_3$,..., $k_{d}$. For example, in $4-$dimensions, $k_2=2, k_3=4, k_4=22$ will make $\phi_c$ independent of $k_1$. Generalization of this result for $d-$dimensional hypercuboids and corresponding proof is given in Appendix A.
 
 For the special case of thin sheets (say $k_2 \rightarrow \infty$ and $k_3 \rightarrow \infty$ with finite $k_1$), we get $\phi_{c}^{k_1} = 1 - \exp\left(-B_c\dfrac{k_1}{\left(8k_1+4\right)}\right)$. For the special case of sticks in 3D (say $k_2 = k_3 = 1$ and finite $k_1$), we get  $\phi_{c}^{k_1} = 1 - \exp\left(-B_c\dfrac{k_1}{5\left(2k_1-1\right)+1}\right)$. Letting the length of the sticks go to infinity, i.e., considering $k_1 \rightarrow \infty$, we get the finite limiting value $\phi_c = 1 - \exp\left(-B_c/10\right)$. As in the case of rectangles on a 2D square lattice, where the limiting case of rectangles with finite width on the lattice doesn't give the continuum model of rectangles, the limiting case of anisotropic shapes in $3D$ cubic lattice also does not give the corresponding continuum model. For example, for cubes and cuboids on a continuum, $\phi_c=1-\exp\left(-B_c/8\right)$ whereas, limiting value obtained for $k_1 \rightarrow \infty$ for sticks ($k_2 = k_3 = 1$) on a cubic lattice is $\phi_c = 1 - \exp\left(-B_c/10\right)$. 
 
 Considering the general case of overlapping and aligned sticks of length $k_1$ randomly placed on a hypercubic lattice in $d$ dimension, we can write the dimension-dependent expression for the excluded volume of sticks, 
\begin{equation}
    V_{ex}=(2d-1)(2k_1-1)+1
\label{eq9}
\end{equation} 
In this case, as the length of the sticks tends to infinity, we get the limiting threshold value in $d$ dimensions as, 
\begin{equation}
    \phi_c = 1-\exp{\left(-B_c\frac{1}{2\left(2d-1\right)}\right)}
\label{eq10}
\end{equation}
Now, for the continuum problem where aligned hypercubes are placed randomly in $d$ dimensional space, we have the percolation threshold $\phi_c = 1-\exp{(-B_c\frac{1}{2^d})}$ \cite{koza}. Assuming $B_c$ to be the same, we have the curious observation that the percolation threshold of sticks whose length tends to infinity and that of hypercubes (squares in 2d) whose sides tend to infinity have the same value in $d \approx 3.66$ dimensions. In other words, for dimensions three and less, the percolation threshold of the infinite hypercubes is larger than that of infinite sticks, whereas for dimensions greater than three, the percolation threshold of infinite sticks is larger than that of infinite hypercubes.

\begin{figure}
    \centering
    \includegraphics[scale=0.57]{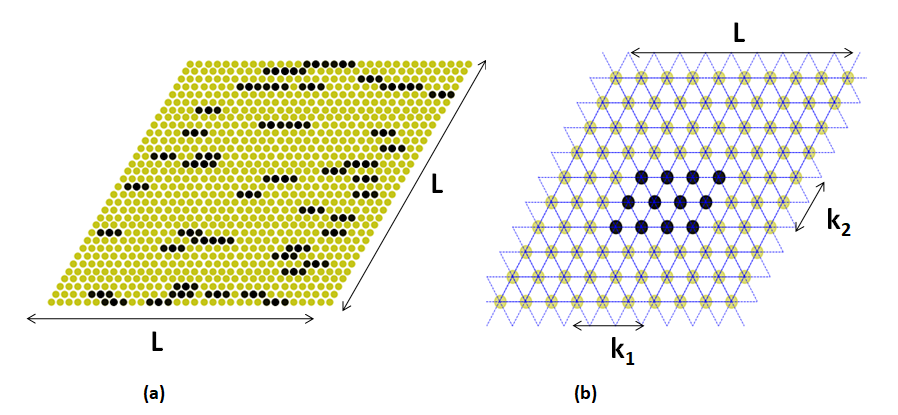}
    \caption{(a) Overlapping trimers (sticks of length $k_1 = 3$ and width $k_2=1$) on a triangular lattice. (b) A parallelogram of $k_1=4$ and $k_2=3$ on a triangular lattice. The excluded volume will be $V_{ex}=\left(2k_1+1\right)\left(2k_2+1\right)-3$.}
    \label{triangularlattice}
\end{figure}

We can see that similar results, as discussed above, will also hold for other lattices. For example, on a triangular lattice, for objects of width $k_2$ and length $k_1$ (a \textquotedblleft parallelogram\textquotedblright), (See Fig.~\ref{triangularlattice}), $V_{ex}=\left(2k_1+1\right)\left(2k_2+1\right)-3$. In this case, we can see that $\phi_c$ is independent of $k_1$ for $k_2 = 1$ (sticks). Considering the limit $k_1 \rightarrow \infty$, we get $\phi_{c}=1-\exp\left(-B_c\dfrac{1}{6}\right)$. Whereas, if we consider parallelograms of equal side-lengths, say $k_1=k_2 = k$, we get $\phi_{c}=1-\exp\left(-B_c\dfrac{1}{4}\right)$  in the limit of $k \rightarrow \infty$.

 The above results show that on $N$-dimensional lattices, when we are considering the percolation of aligned overlapping shapes that can be described by $M$ independent length parameters, the following holds. a) For a specific choice of values for $(M-1)$ of these parameters, the percolation threshold can become independent of the remaining parameter. b) Whenever we let a subset of the $M$ parameters go to infinity while keeping the remaining parameters finite, we expect a finite percolation threshold, which depends upon the finite parameter values. c) These limiting threshold values, in general, will differ from the corresponding continuum results. 

We can perform the excluded volume theory calculations for other shapes in Fig.~\ref{figure1}(c). For diamonds of size $k$ (See Fig.~\ref{figure1}(c)), the area of the diamond is $V = \left(\frac{k^2+1}{2}\right)$ where $k$ must be odd. If we choose the central site of the diamond as the index site, the shape of the excluded area will be a bigger diamond of size $\left(2k+1\right)$. We can calculate the areas and excluded areas of isosceles triangles, and hexagons in a similar way. If we choose the central site of the base of the isosceles triangle as the index site, the shape of the excluded area will be a hexagon of dimensions $\left(2 k +1\right)$ and $\left(k+1\right)/2$. For the hexagon, the shape of the excluded area will also be a hexagon of dimensions $\left(2k_1+1\right)$ and $\left(2k_2+1\right)$. 

In table \ref{table1}, we provide a summary of the expressions for the percolation thresholds obtained for a few different shapes using the discrete version of excluded volume theory. Wherever possible, we have included numerical values for the limiting cases, where $B_{c}$ values for different shapes are taken from existing studies of the continuum percolation problem of the corresponding shapes. Note that the diamond shape on an upright square lattice is equivalent to a square shape on a diagonal square lattice \cite{johnston}. Hence, we expect that the $B_c$ values for the two are the same. In fact, Gouker and Family~\cite{gouker} have studied the percolation of diamond shapes on a square lattice. In Table \ref{table2}, we compare the theoretical predictions with the simulation values given in ~\cite{gouker}. We can see an excellent agreement between the two at larger diamond sizes.

The following section presents simulation results for aligned and overlapping rectangles that closely agree with the predicted behavior. In particular, the agreement between the theory and simulation results seems exact for the limiting cases.
\setlength{\extrarowheight}{15pt}
\begin{table*}[!ht]
\centering
\begin{tabular}{ |p{3cm}|p{2cm}|p{7cm}|p{4cm}|}
\hline
Shape & Lattice  & CCVF $\phi_{c}$ from discrete excluded volume theory & Limiting Values of $\phi_{c}$ \\
\hline
\hline
Rectangles of size $k_1 \times k_2$ & 2D Square & $1-\exp\left(-B_{c}\frac{k_{1} k_{2}}{(2k_{1}+1)(2k_{2}+1)-5}\right)$ & $1-\exp\left(-B_{c}\frac{k_{2}}{4k_{2}+2}\right)$ ($k_{1}\rightarrow\infty$, finite $k_{2}$)\\
\hline
Squares ($k_1 = k_2 = k$) & 2D Square &  $1-\exp\left(-B_{c}\frac{k^{2}}{(2k+1)^{2}-5}\right)$ & $0.6667$ ($k\rightarrow \infty$), $B_c=4.3953711(5)$\cite{koza}\\
\hline
Diamonds of linear size $k$ & $2D$ Square & $1-\exp\left(-B_{c}\frac{k^{2}+1}{4k(k+1)}\right)$ & $1-\exp\left(-B_{c}/4\right)$ ($k \rightarrow \infty$)\\
\hline
Triangles (isosceles of base $k$)& 2D Square & $1-\exp\left(-B_{c}\frac{k^2+2k+1}{2\left(3k^2+6k-1\right)}\right)$ & $1-\exp\left(-B_{c}/6\right)$ ($k \rightarrow \infty$)\\
\hline 
Hexagon with dimensions $k_1$ and $k_2$ &  2D Square & $1-\exp\left(-B_{c}\frac{k_1+2k_2(k_1-k_2-1)}{(2k_1+1)+2(2k_2+1)(2k_1-2k_2-1)-1}\right)$& $1-\exp\left(-B_{c}\frac{k_{2}}{4k_{2}+2}\right)$ ($k_1 \rightarrow \infty$, finite $k_2$)\\
\hline
Cubes ($k_1 = k_2 = k_3 = k$) & 3D Cubic & $1-\exp\left(-B_{c}\frac{k^{3}}{(2k-1)^{3}+6(2k-1)^{2}-1}\right)$&$0.2773$ ($k\rightarrow \infty$), $B_c=2.5978(5)$\cite{koza}\\
\hline
Sticks of length $k_1$ & 3D Cubic & $1-\exp\left(-B_{c}\frac{k_1}{5(2k_1-1)+1}\right)$ & $0.22877$ ($k_1\rightarrow \infty$) , $B_c=2.5978(5)$\cite{koza} \\
\hline
Parallelograms of size $k_1 \times k_2$  & 2D Triangular & $1-\exp\left(-B_{c}\frac{k_1 k_2}{(2k_1+1)(2k_2+1)-3}\right)$ & $1- \exp\left(-B_{c}\frac{k_{2}}{4k_{2}+2}\right)$ ($k_1\rightarrow \infty$, finite $k_2$) \\
\hline
\end{tabular}
\caption{Expressions for CCVF $\phi_{c}$ for various shapes and lattices from discrete excluded volume theory. For obtaining the numerical values in the last column, the $B_{c} $ value for a shape is assumed to be that of the continuum percolation problem of the same shape.}
\label{table1}
\end{table*}

\begin{table*}[!ht]
\centering
\begin{ruledtabular}
\centering
\begin{tabular}{ccccc}
$k$ & $\phi_c$ using the expression in the third row of Table.~\ref{table1} & $z$ & $p_c$ & Simulation results for $p_c$ from previous studies~\cite{gouker}.\\
\hline
5& 0.61416 & 12 & 0.3174& $0.290\pm0.005$\\
9& 0.63255 & 40 & 0.100117& $0.105\pm0.005$\\
13 & 0.6417 & 84& 0.04888& $0.049\pm0.005$\\
17& 0.6470& 144&0.028927&$0.028\pm0.005$\\
21 & 0.6505&220&0.01911&$0.019\pm0.005$\\

\end{tabular}
\end{ruledtabular}
\caption{Comparison of discrete excluded volume theory results for diamond shapes on a square lattice with previous simulation results \cite{gouker}. $p_c$ is the critical number of objects per site and is related to $\phi_c = 1 - \exp\left(-\eta_c\right)$ via the relation $p_c =  \dfrac{4 \eta_c}{z}$~\cite{xun}. Interaction range $R$ in \cite{gouker} is related to $k$ by $k=2R+1$. $z$ is the coordination number \cite{gouker}.}
\label{table2}
\end{table*}

\subsection{Simulation results}
\label{subsec2}
\begin{figure*}[hbtp]
\includegraphics[scale=0.56]{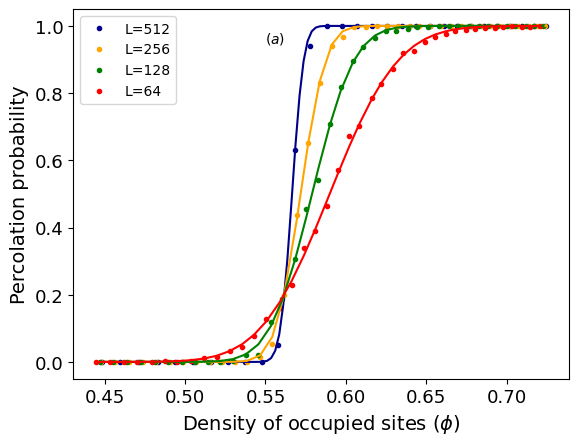}\hfill
\includegraphics[scale=0.56]{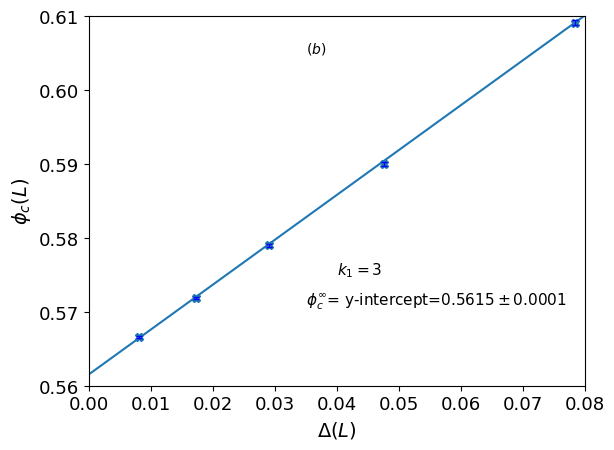}
\caption{(a) Variation of the percolation probability with the density of occupied sites $\phi$ for $k_1=3, k_2 = 1$ rectangles for different system sizes $L$. {color{red} Solid lines are best-fit functions (see text).}(b) Variation of the corresponding effective percolation threshold $\phi_{c}^{L}$ with the width of the transition region $\Delta(L)$ along with the best straight-line fit. $Y-$intercept of the graph yields the threshold in the limit of infinite system size.}
\label{figure3}
\end{figure*}

We simulate the model of aligned and overlapping rectangles on a two-dimensional square lattice and study its percolation properties. To construct the model, aligned rectangles of size $k_1 \times k_2$ are distributed uniformly and randomly on a square lattice of size $L \times L$.  The percolation probability is determined by detecting the presence of a spanning cluster in the vertical direction using the standard Hoshen-Kopelman algorithm \cite{hoshen}.  Later in Sec. \ref{subsubsec1}, we verify that the results remain unchanged if we consider the horizontal direction for defining the spanning cluster. For each system size and density of occupied sites, we generate a number of samples, and the percolation probability is evaluated as the fraction of samples that are percolating. The number of samples considered varies between $10^3$ for $L \leq 256 $ and $10^2$ for higher values of $L$. Percolation probability is plotted against the density of occupied sites for different system sizes $L$, and by fitting each curve with the function $\frac{1+\textrm{erf}{[\phi-\phi_{c}(L)/\Delta(L)]}}{2}$, we obtain the effective percolation threshold $\phi_{c}(L)$ for system size $L$ and width of the transition region $\Delta(L)$ \cite{rintoul}. We have the linear scaling relation,
\begin{equation}
     \phi_{c}(L) = B*\Delta(L) + \phi_{c}(\infty) 
\label{eq11}
\end{equation}
where $\phi_{c}(\infty)$ is the percolation threshold in the limit of infinite system size $L\rightarrow \infty$, and $B$ is a constant \cite{stauffer}. 
A typical example of the plot of percolation probability against the density of occupied sites for $k_1 = 3, k_2 = 1$ rectangles for different values of $L$ and the corresponding plot of $\phi_{c}(L)$ against $\Delta (L)$ is shown in Fig. \ref{figure3}. $Y-$intercept of the latter plot gives the percolation threshold in the $L \rightarrow \infty$ limit.
The percolation threshold is determined for rectangles of widths $k_2 = 1, 2$ and $3$ for increasing values of $k_1$, and is shown in Table \ref{table3} along with the values based on the lattice version of excluded volume theory discussed in Sec.\ref{subsec1}. For the value of $B_c$, we take that obtained from continuum simulations of aligned squares or rectangles ($B_{c}=4.3953711(5)$ \cite{koza}) to get numerical estimates of percolation thresholds.  


\begin{table*}[!ht]
\centering
\begin{ruledtabular}
\centering
\begin{tabular}{cccccc}

Length $k_1$  & $\phi_c^{k_1}$, $k_2=1$ & $\phi_c^{k_1}$, $k_2 = 1$ Theory & $\phi_c^{k_1}$, $k_2=2$ & $\phi_c^{k_1}$, $k_2 = 3$ & $\phi_c^{k_1}$, $k_2 = 3$ Theory  \\
\hline

1 & $0.5927(4)$ &$0.6667$ & - & - &-\\

2&$0.5715(18)$ &$0.5848$ &$0.5837(2)$ & - &-\\

 3 & $0.5615(1)$&$0.5614$ &$0.5864(5)$ &$0.5948(6)$ &$0.5930$\\

4& $0.5523(9)$ &$0.5503$ & $0.5903(5)$ & $0.5965(4)$& $0.5972$\\

5& $0.5497(19)$ &$0.5438$ & $0.5916(5)$ &$0.597(2)$ &$0.5997$\\

7& $0.5422(6)$ &$0.5366$ &$0.5909(26)$ &$0.603(3)$ &$0.6027$\\

9& $0.5402(46)$ & $0.5326$&$0.5933(8)$ &$0.6048(7)$&$0.6043$ \\

11&$0.540(1)$ & $0.5302$ &$0.5922(27)$ &$0.605(2)$ &$0.6053$\\

13& $0.539(4)$ & $0.5285$&$0.5950(16)$ &$0.606(2)$ &$0.6061$\\

15& $0.538(4)$ &$0.5272$ &$0.592(4)$ & $0.608(2)$ &$0.6066$\\

17& $0.536(4)$ &$0.5263$ &$0.593(7)$ &$0.609(2)$ &$0.6070$\\

 19& $0.529(6)$ & $0.5255$&$0.5927(41)$ & $0.609(1)$ &$0.6073$\\

21& $0.528(7)$ &$0.5249$ &$0.5928(14)$ &$0.6102(1)$ &$0.6076$\\
\end{tabular}
\end{ruledtabular}
\caption{\small Percolation thresholds of aligned and overlapping rectangles of length $k_1$ and width $k_2=1, 2$ and $3$ on a square lattice determined from simulations. The corresponding results from the discrete excluded volume theory for $k_2 = 2$ and $3$ are also given for comparison. For $k_2 = 2$, the theory gives a constant value for $\phi_c^{k_1} = 0.5848$.}
\label{table3}
\end{table*}

\begin{figure}[hbtp]
\centering
\includegraphics[scale=0.55]{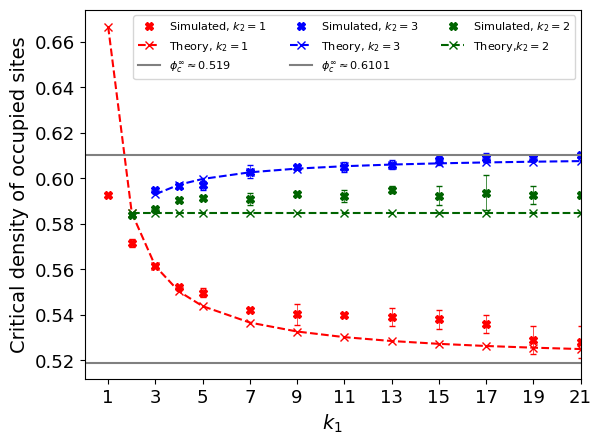}
\caption{Percolation threshold $\phi_{c}$ vs $k_1$ for rectangles of width $k_2 = 1, 2$, and $3$. The dashed lines represent the corresponding predicted values based on the discrete excluded volume theory, and the solid horizontal lines represent the thresholds in the limit $k_1 \rightarrow \infty$ obtained from Equation~(\ref{eq5}).}
\label{figure4}
\end{figure}

Plots of the percolation threshold against $k_1$ for rectangles of width $k_2 = 1, 2$ and $3$ are shown in Fig. \ref{figure4}, confirming the trends predicted by the theory. As predicted by the excluded volume theory, for $k_2 = 1$, the CCVF is monotonically decreasing, whereas for $k_2 = 3$, it is monotonically increasing with $k_1$. For $k_2=2$, critical density is nearly a constant.  Percolation thresholds for the limiting case of $k_1 \rightarrow \infty$ are also shown in the figure.

\subsubsection{Isotropy of percolation threshold}
\label{subsubsec1}
\begin{figure}[hbtp]
\centering
\includegraphics[scale=0.55]{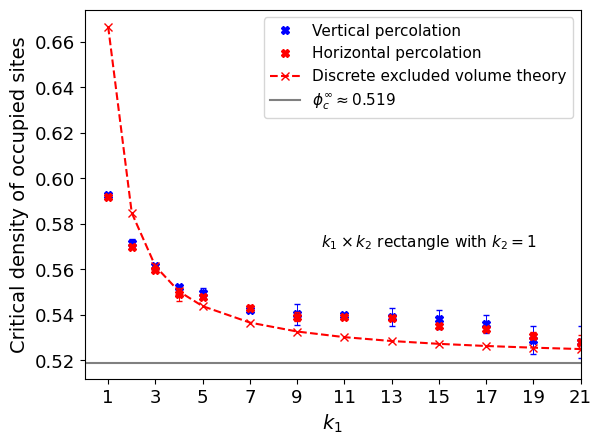}
\caption{Percolation threshold versus stick length $k_1$ for two different definitions of percolation. In one case, percolation is defined as the emergence of a spanning cluster in the vertical direction (perpendicular to the direction of alignment of sticks). In the other scenario, it is defined as the emergence of a spanning cluster in the horizontal direction (in the direction of alignment of sticks). Results from the excluded volume theory and the limiting value of the threshold for sticks of infinite length are also shown.}
\label{figure5}
\end{figure}

When considering $k_1 \times k_2$ rectangles on the square lattice, we fixed $k_2$ and varied $k_1$. For $k_1 > k_2$, this means that there is a preferred direction for the percolation to happen in finite systems. i.e., along the horizontal direction. In the last section, we defined the system as percolating when there is a spanning cluster in a direction perpendicular to the direction of alignment. A natural question is whether the percolation point is different if we define a configuration as percolating when there is a spanning cluster in the direction of alignment. In continuum percolation models, previous studies suggest isotropy of the percolation threshold even for highly anisotropic systems like rectangles with large aspect ratios \cite{klatt}. It is found that the effective percolation threshold for a finite system size in the direction of alignment will always be smaller compared to that in the orthogonal direction. However, this difference will decrease with increasing system size and finally vanish in the infinite system size limit. We verify that this isotropy of the percolation threshold holds for aligned rectangles on the square lattice as well. The critical density of occupied sites evaluated (See Fig. \ref{figure5}) shows that the percolation threshold is also isotropic in the case of lattices. Just as in the continuum case, this isotropy may be explained based on the uniqueness of the spanning cluster in percolation problems \cite{klatt}.

\subsubsection{Critical exponents for aligned stick model}
\label{subsubsec2}
\begin{figure*}[hbtp]
\includegraphics[scale=1]{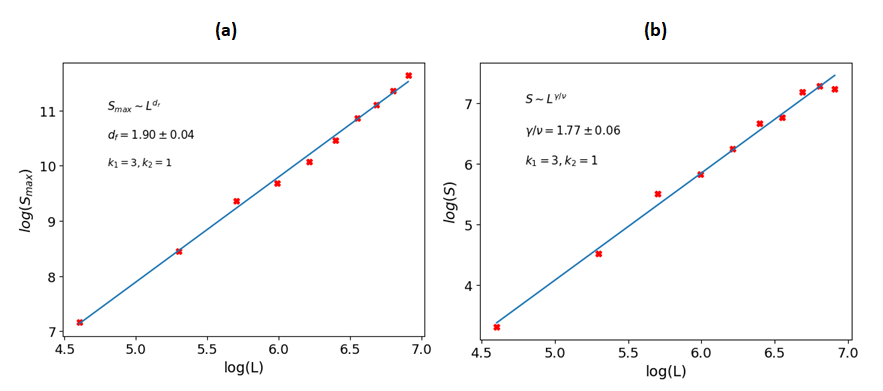}
\caption{(a) Variation of the size of the largest cluster $S_{max}$ with system size $L$ along with the best straight line fit for $k_1=3, k_2 = 1$. The slope of the log-log plot gives the fractal dimension $d_{f}$. (b) Variation of the size of the finite clusters $S$ with system size $L$ along with the best straight line fit. The slope of the log-log plot gives the ratio of critical exponents $\frac{\gamma}{\nu}$.}
\label{figure6}
\end{figure*}

\begin{figure*}[hbtp]
\includegraphics[scale=1]{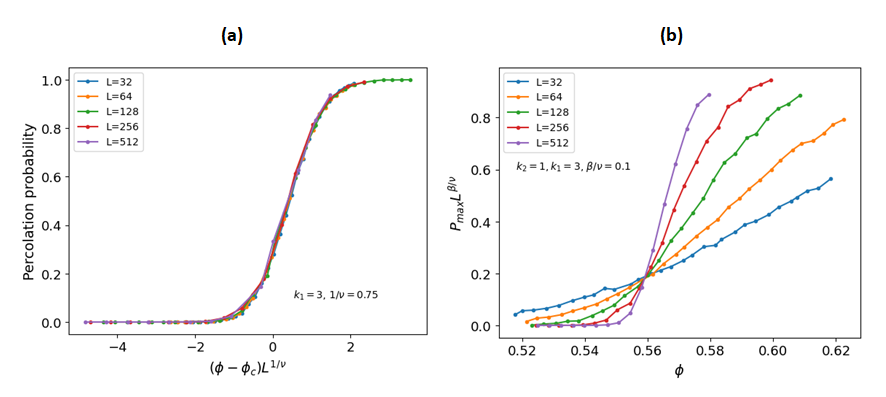}
\caption{(a)Plots of percolation probability against $(\phi-\phi_{c})L^{\frac{1}{\nu}}$ for $k_1=3, k_2 = 1$ with $\frac{1}{\nu}=0.75$ and  $\phi_c = 0.5615$.
(b) Plot of $P_{max}L^{\beta/\nu}$ vs $\phi$ for $k_1=3$ and $k_2=1$ for $\frac{\beta}{\nu}=0.1$, $\frac{1}{\nu}=0.75$}
\label{figure7}
\end{figure*}

Critical exponents are evaluated using finite size scaling methods \cite{christensen,stauffer}. If $S_{max}$ is the size of the largest cluster and $S$ is the total size of the remaining clusters, we have the scaling relations,
\begin{eqnarray}
S_{max} &\sim L^{d_{f}}\\
S &\sim L^{\frac{\gamma}{\nu}}
\end{eqnarray}
where $d_f$ is the fractal dimension, $\gamma$ is the exponent corresponding to mean cluster size and $\nu$ is the correlation length exponent. The plots of  $S_{max}$ against system size $L$ and $S$ against $L$ for $k_1 = 3, k_2=1$ are shown in Fig.~\ref{figure6}(a) \& (b) respectively. The slopes of log-log plots give the fractal dimension $d_{f} = 1.90\pm0.04$ and the ratio $\frac{\gamma}{\nu} = 1.77\pm0.06$. 
The percolation probability $P(\phi, L)$ is expected to scale with the system size $L$ as 
\begin{equation}
P(\phi)= f((\phi-\phi_{c})L^{\frac{1}{\nu}})
\label{eq14}
\end{equation}
where $f$ is the scaling function. This implies that the curves of percolation probability for different $L$ when plotted as a function of $(\phi-\phi_{c})L^{\frac{1}{\nu}}$ with the correct value for $\phi_c$ and $\nu$, will collapse to a single curve. In Fig. \ref{figure7}(a), we verify that curves for various system sizes fall on top of each other for $\frac{1}{\nu}=0.75$ and $\phi_c = 0.5615$. 

The other critical exponents are evaluated by making use of the relations, 
\begin{equation}
d_{f} = d-\frac{\beta}{\nu}
\end{equation} 
\begin{equation}
\sigma=\frac{1}{\nu d_{f}}
\end{equation}
\begin{equation}
\tau = 1+\frac{d}{d_{f}}
\end{equation}

We obtain $\nu \approx 1.33$, $\beta \approx 0.133$, $\sigma \approx 0.3948$, $\gamma \approx 2.36$ and $\tau \approx 2.05$. 
Finally, if $P_{max}(\phi)$ is the probability of a site belonging to the largest cluster, it is expected to scale with the system size $L$ as,
\begin{equation}
P_{max}(\phi)= L^{-\frac{\beta}{\nu}}F((\phi-\phi_{c})L^{\frac{1}{\nu}})
\end{equation}
where $F$ is the scaling function and $\beta$ is the order parameter exponent. When $P_{max}L^{\frac{\beta}{\nu}}$ is plotted against $\phi$ for different $L$, all curves will pass through a single point which is at $\phi=\phi_c$ as verified in Fig.~\ref{figure7}(b).

The obtained values of various critical exponents align with that of the standard two-dimensional percolation problem which are $\beta=5/36$, $\gamma=43/18$, $\nu=4/3$, $\sigma=36/91$, $\tau=187/91$ and $d_f=91/48$ \cite{christensen}. Similar values are obtained for other values of $k_1$ and $k_2$ as well. Hence, we can conclude that the percolation problem of aligned overlapping rectangles belongs to the standard percolation universality class. Note that the above results were obtained only for a limited range of $k_1$ and $k_2$ values. There remains the possibility that the value of the critical exponents could be different for higher values of $k_1$ and $k_2$, especially in the limiting cases of $k_1 \rightarrow \infty$ with finite $k_2$.

\subsubsection{Comparison with rectangles of mixed orientation}
\label{subsubsec3}
\begin{figure}[hbtp]
\includegraphics[scale=0.6]{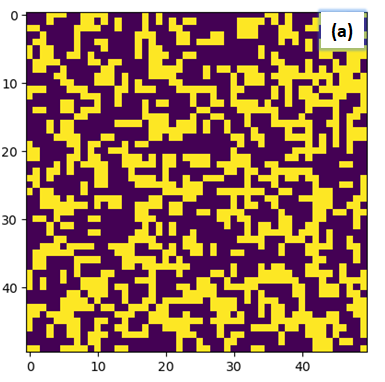}
\includegraphics[scale=0.6]{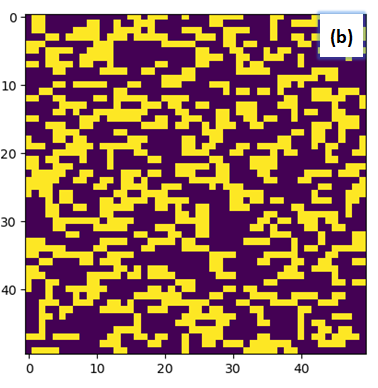}
\caption{(a) Horizontal and vertical dimers on $50 \times 50$ square lattice with the density 
of occupied sites 0.5. $s\approx 0$. (b)Horizontal and vertical dimers on $50 \times 50$ 
square lattice with $s\approx 0.66$}
\label{figure8b}
\end{figure}

We consider the percolation of overlapping sticks having both orientations (rectangles with $k_1 = 1$ or $k_2 = 1$ in our notation) here. The problem has been considered earlier in \cite{mecke}. To study the percolation as we vary the relative fraction of sticks having horizontal and vertical orientations, we can define the parameter \cite{tarasevich1}
\begin{equation}
s=\frac{|\left(n_{v}-n_{h}\right)|}{|\left(n_{v}+n_{h}\right)|}
\label{eq19}
\end{equation}
where $n_v$ and $n_h$ represent the fraction of sticks that are vertically and horizontally oriented, respectively (See Fig.~\ref{figure8b}). Thus, $s=1$ corresponds to the fully aligned case, and $s=0$ corresponds to the isotropic case where the average fraction of sticks that are vertically and horizontally oriented is equal. The $s=0$ case has been considered earlier in~\cite{mecke}. The values of the percolation threshold obtained for a few different values of the parameter $s$ are given in Table~\ref{table4} along with the earlier results for $s=0$ case. We can see that the percolation threshold increases with an increase in the degree of alignment, which is also seen in earlier results for non-overlapping sticks \cite{tarasevich1}.

\begin{table*}[!ht]
\centering
\begin{ruledtabular}
\centering
\begin{tabular}{ccccc}
$s$ & Length of  the sticks & $\phi_c$ & $\phi_c$ for $s=0$ from \cite{mecke}&  $\phi_c$ from theory (See Appendix B)\\
\hline
$s=0$& 2& $0.5483(2)$ & $0.54691$ &  $0.5503$\\
$s=0.66$ & 2 & $ 0.55885(14)$ & & $0.5651$\\
$s=1$& 2 & $0.5715(18)$ & & $0.5848$\\
$s=0$& 3& $0.50004(64)$ & $0.49898$ & $0.5097$\\
$s=0.33$ & 3 & $ 0.5325(38)$ & & $0.5150$\\
$s=1$ & 3 & $0.5615(1)$ & &$0.5614$\\
$s=0$ & 4 & $0.4562(7)$ & $0.45761$ & $0.4785$\\
$s=0.33$ & 4& $0.4665(2)$ & & $0.4855$\\
$s=0.66$ &4& $0.4944(17)$ & &$0.5081$\\
$s=1$ & 4 & $0.5523(9)$ & & $0.5503$\\

\end{tabular}

\end{ruledtabular}
\caption{Effect of alignment in dimer ($k=2$ stick) and trimer ($k=3$ stick)percolation. The critical density of occupied sites $\phi_c$ for different values of $s$ is shown. The values of $\phi_c$ for the special case of $s=0$ from \cite{mecke} are also given. Note that in \cite{mecke}, the thresholds are given in terms of $p_c$, which is the critical number of objects per site. $\phi_c$ is related to $p_c$ via the relation $\phi_c=1-\left(1-p_c\right)^{2k}$.  Theoretical values of $\phi_c$ calculated from the $s-$dependent excluded volume theory relation (See Appendix B) assuming $B_c\approx 4.3953711$ is given for comparison.}
\label{table4}
\end{table*}

 We can derive an $s-$ dependent expression for the excluded area as well as $\phi_c$ for overlapping rectangles of mixed orientations on a square lattice. Details of the derivation and results are given in Appendix B. The results confirm that the interesting behavior observed for the thresholds for fully aligned rectangles is unique, and we don't expect it for values of $s$ different from 1 (fully aligned case). For $s<1$, for large $k_1$, the percolation threshold is found to be a monotonically decreasing function of $k_1$. Theoretical values obtained from the $s-$ dependent expression for $\phi_c$ in Appendix B is also given in Table \ref{table4} for comparison. Even with the low values of length considered, the deviation from simulation results is typically a few percentage, with the best accuracy seen for the fully aligned case ($s=1$) where the deviation is typically less than 1\%.


\subsubsection{Discrete to continuum transition}
\label{subsubsec4}
Models of extended shapes on lattices interpolate between discrete and continuum percolation models. With symmetric shapes like squares, letting the size of the squares go to infinity - which is equivalent to letting the lattice spacing go to zero - for an infinite lattice - yields the continuum limit \cite{koza1}. The convergence of discrete hypercubes and hyperspheres to corresponding continuum models and the universality of the convergence rate is discussed in \cite{koza1,brzeski}. However, with anisotropic shapes like rectangles considered here, letting the length of one side of the shape go to infinity while keeping the other side fixed  - similar to letting the lattice spacing go to zero only along one direction - does not seem to correspond to any usual continuum percolation problem. Note that the continuum percolation threshold for aligned rectangles of all aspect ratios is the same as that of aligned squares \cite{klatt}. However, as we saw, no corresponding result exists for lattices. Just to emphasize the difference in values, we find the limiting value of $2D$ stick percolation model as $\phi_{c}^{\infty} \approx 0.519$, whereas the continuum percolation threshold for aligned rectangles is  $\phi_{c} \approx 0.667$ \cite{klatt}. Similarly, in three dimensions, we find the limiting value of the $3D$ stick model as $\phi_{c} \approx 0.229$, whereas the continuum percolation threshold value for overlapping cubes is $\phi_{c} \approx 0.277$ \cite{baker}.

\begin{figure}
\centering
\includegraphics[scale=0.6]{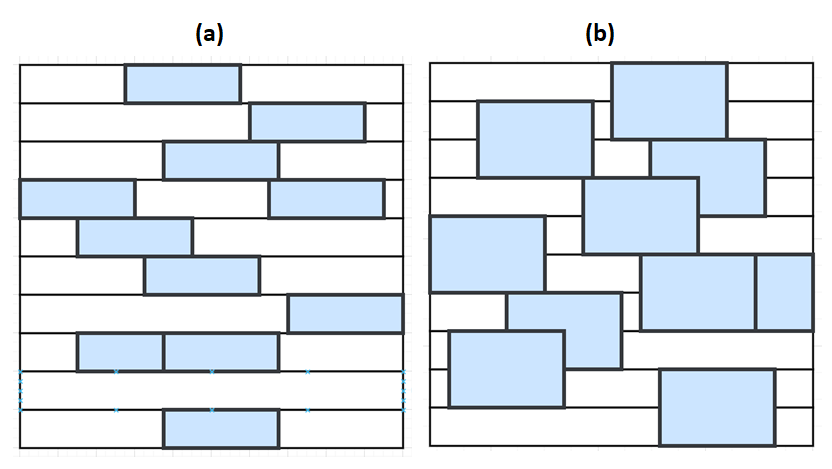}
\caption{Schematic representation of percolation models of aligned, overlapping rectangles on a 2D semi-continuum geometry. Rectangles whose widths are restricted to integer multiples of lane width are dropped into lanes with a specific number density. Two rectangles are considered to be connected if they overlap or share edges. (a) Rectangles with width $k_2=1$. (b) Rectangles with width $k_2=2$.}
\label{semicontinuum}
\end{figure}

The appropriate \textquoteleft continuum\textquoteright\; analogue of the case where we let the length of the rectangles on the lattice go to infinity while keeping its width finite can be seen as a semi-continuum system in which rectangles are dropped into parallel lanes. The rectangles can be placed anywhere in each lane (horizontal direction), but they cannot straddle between the lanes. A schematic representation of the system for two different widths of the rectangles is shown in Fig.~\ref{semicontinuum}. 

\section{CONCLUSIONS}
\label{sec3}
Percolation of overlapping and extended shapes on lattices that interpolate between discrete and continuum models is a topic of recent interest. This work investigates the percolation of aligned, overlapping, and anisotropic shapes on lattices. First, we use a lattice version of excluded volume theory to show that shape-anisotropy leads to some intriguing consequences regarding the percolation behavior. As a quintessential example of an anisotropic object, we investigate the case of aligned and overlapping rectangles on a square lattice in detail and show that for rectangles of width unity (sticks), the percolation threshold is a monotonically decreasing function of the stick length. In contrast, for rectangles of width greater than two, the percolation threshold is a monotonically increasing function of the length. The percolation threshold is independent of its length for rectangles of width two. This independence of threshold on the length of a side holds for $d-$dimensional hypercubiods as well for specific integer values for the lengths of the remaining sides. The limiting case of the length of the rectangles going to infinity shows some remarkable behavior. It is found that the limiting value depends upon the rectangle's width. This is in stark contrast to the continuum percolation of aligned overlapping rectangles for which the percolation threshold is independent of the aspect ratio. The general notion that continuum percolation of aligned objects can be considered the limiting case of corresponding discrete model \cite{koza1} is only valid for symmetric shapes. Our research extends beyond rectangles, demonstrating that similar results hold for anisotropic objects on various lattices and dimensions and give trends and numerical estimates of percolation thresholds for several such shapes and lattices. 

We present simulation results for the percolation thresholds, which verify the predictions of the excluded volume theory. The percolation thresholds from simulations show that the results of the lattice-excluded volume theory give very good results, and its accuracy increases with the size of the extended shapes. 
We also determine the critical exponents for the model using simulations, which confirm that the model with finite rectangles belongs to the same universality class as the usual lattice percolation. 

In addition, we verify the isotropy of the percolation threshold by obtaining the thresholds for percolation transition with the spanning cluster considered in the direction of alignment and the orthogonal direction. We also compare our results with models where rectangles of mixed orientation are allowed and models where non-overlapping sticks with varying degrees of anisotropy are considered. Our simulation results show that alignment increases the percolation threshold even for overlapping shapes.

The findings of this study contribute to a deeper understanding of percolation behavior in lattice systems with overlapping shapes and shed light on the intricate interplay between discrete and continuum models. Our results show that the lattice version of the excluded volume theory is remarkably accurate in describing the behavior of percolation of aligned and overlapping shapes on lattices qualitatively and quantitatively. 

    Our study opens up many avenues for further research. In this work, for obtaining the numerical results for the thresholds using the discrete excluded volume theory, we used the value of $B_c$ - the average number of connections an object has at the critical point - from continuum percolation. This, essentially assumes that the connectivity properties of the lattice system of overlapping shapes are the same as that of the corresponding continuum system at the critical point. In other words, we are assuming that the connectivity network formed at critical points by discrete systems like the one we considered here is similar to continuum critical network of similar shapes. It is desirable to obtain the value of $B_c$ based on considerations of lattice systems itself. Methods like the ones discussed in \cite{fabian,fabian1,fabian2,mecke1,mecke2} may provide alternate ways of looking into this problem.

Other factors like polydispersity of anisotropic shapes will affect the critical behavior of the percolation problem discussed \cite{tarasevich}. Especially the effect of unbounded size distributions on the critical behavior is a problem of interest \cite{sasidevan,hall}. Another potential area for further investigation is the relation to extended neighborhood percolation models \cite{xun}. The \textquoteleft continuum\textquoteright\; limiting case of the lattice percolation models considered here can be seen as percolation models on semi-continuum geometries (Fig.~\ref{semicontinuum}) where basic percolating units can occupy only positions in lanes, which is another interesting problem worth investigating. This novel semi-continuum model of percolation may have relevance in describing random geometry of materials with a layered structure, such as layered rocks \cite{bai,xu}. A detailed description and analysis of the model will be attempted in future work.\\

\begin{acknowledgments}
VS acknowledges support from the University
Grants Commission-BSR Start-up Grant No:F.30-415/2018(BSR). The authors would like to thank Robert M Ziff for useful comments on an earlier version of the manuscript.
\end{acknowledgments}

\appendix
\section{Independence of percolation threshold $\phi_c$ on  side length $k_1$ for a particular set of integer values of side lengths $k_2, k_3, k_4, .....k_d$ for aligned hypercuboids on a $d-$ dimensional hypercubic lattice}
Consider hypercuboids with side lengths $k_1,k_2,k_3....,k_d$ on a $d-$ dimensional hypercubic lattice. Hypercuboids are randomly distributed and aligned along the direction of $k_1$. In Sec.~\ref{subsec1}, we showed that in 3-dimensions, there is a unique set of integer values of $k_2$ and $k_3$, which will make the percolation threshold independent of $k_1$. Here, we generalize this result to $d-$dimensions. We show that there is always a set of integer values of $k_2,k_3,.....,k_d$ such that the threshold is independent of $k_1$. 

Consider $(d-1)$ dimensions, assume that we have a set of integers $k_2, k_3,..., k_{d-1}$ which makes $\phi_c$ independent of $k_1$. Let $a_i=(2k_i-1)$ for $i=1,2,....,(d-1)$. Then generalizing Equation~(\ref{eq7}), the expression for the excluded volume of aligned hypercuboids in $(d-1)$ dimensions can be written as

\begin{multline}
V_{ex}=a_1 A_1+ 2\bigl[a_1 A_{d-1}+a_1 A_{d-2}+a_1 A_{d-3}+......+\\....a_1 A_4+a_1 A_3+a_1 A_2+A_1\bigr]-1.
\label{equation1}
\end{multline}

where, 
\begin{equation*}
    A_1=\prod_{i=2}^{(d-1)} a_i
\end{equation*} 
and
\begin{equation*}
    A_j=\prod_{i=2, i \neq j}^{(d-1)} a_i
\end{equation*}

For $\phi_c$ to be independent of $k_1$, the ratio $\frac{V}{V_{ex}}$ shouldn't contain any $k_1$ in it, which means that $V_{ex}$ has to be a multiple of $k_1$. For this to happen, for a particular choice of integers $k_2, k_3,...k_{d-1}$, we will have an expression of the form $V_{ex} = B_1 a_1 +B_2$  where $B_1$ and $B_2$ are constants for a particular choice of $a_2, a_3,...a_{d-1}$. Now $V_{ex}$ will be a multiple of $k_1$ if and only if $B_1=B_2$. Identifying $B_1$ and $B_2$ from Equation~(\ref{equation1}) leads to the condition,

\begin{multline}
2\bigl[A_{d-1}+A_{d-2}+A_{d-3}+...\\...+A_4+A_3+A_2\bigr]=A_1-1
\label{equation2}
\end{multline}

Now assume that Equation~(\ref{equation2}) is satisfied in $(d-1)$ dimensions for a particular set of integer values $a_1,a_2,.....,a_{d-1}$. Using these values in the $d-$dimensional problem, we can write the expression for $V_{ex}$ in $d-$dimensions as,
\begin{multline}
V_{ex}=a_1A_1a_{d}+ 2\bigl[a_1A_1+a_1A_{d-1}a_{d}+a_1A_{d-2}a_{d}+....\\........+a_1A_{4}a_{d}+a_1A_{3}a_{d}+a_1A_{2}a_{d}+A_1a_{d}\bigr]-1
\label{equation3}
\end{multline}
where $a_d$ is still to be determined.

For $\phi_c$ to be independent of $k_1$, we now have the condition,

\begin{multline}
2\bigl[A_{1}+A_{d-1}a_d+A_{d-2}a_d+.....\\.........+A_{4}a_d+A_{3}a_d+A_{2}a_d\bigr]\\=A_1a_{d}-1
\label{equation4}
\end{multline}

From this, we get 
\begin{equation}
a_d=\frac{-\left(1+2A_1\right)}{\left[2\left(A_{d-1}+A_{d-2}+A_{d-3}+.....+A_{3}+A_{2}\right)-A_1\right]}
\label{equation5}
\end{equation}

Now using Equation~(\ref{equation2}) in this, we get
\begin{equation}
a_{d}=1+2A_1
\label{equation6}
\end{equation}

Since $A_1$ is a product of integers, $a_d$ is always an odd positive integer. Since Equation~(\ref{equation2}) is satisfied in 3 dimensions, we can conclude that there is always a set of integer values for $k_2,k_3,.....,k_d$, which will make the percolation threshold of hypercuboids on a $d-$dimensional cubic lattice independent of $k_1$.   Solutions obtained for a few dimensions are listed in Table~\ref{table5}.

\begin{table*}[t]
\begin{ruledtabular}
\begin{tabular}{ccc}
Dimension $d$ &Values of $a_2,a_3,....,a_{d}$& Set of integer solutions $k_2,k_3,....,k_d$ \\
\hline
2& 3 & 2\\
3& 3, 7 & 2, 4\\
4& 3, 7, 43 & 2, 4, 22\\
5& 3, 7, 43, 1807 & 2, 4, 22, 904\\
6& 3, 7, 43, 1807, 3263443 & 2, 4, 22, 904, 1631722\\

\hline
\end{tabular}
\end{ruledtabular}
\caption{Integer solutions $a_i$ and corresponding $k_i$ values obtained for various dimensions $d$ using Equation~\ref{equation6}.}
\label{table5}
\end{table*}

\section{Excluded volume results for multi-oriented rectangles on a square lattice}

Consider a square lattice of size $L\times L$. $N$ units of percolating objects are placed randomly. Each unit of percolating objects is a rectangle with side lengths $k_1$ and $k_2$.  Assume that $k_1 > k_2$. Assume that out of total $N$, $n_{v}$ rectangles are placed vertically (i.e, side with size $k_1$ in the $Y$ direction) and $n_{H}$ are placed horizontally (i.e, side with size $k_1$ in the $X$ direction). As in Sec.\ref{subsubsec3}, we define a parameter $s=\frac{|n_{v}-n_{h}|}{n_{v}+n_{h}}$.  Alternatively, if we define $p_h$ as the probability of a rectangle being placed horizontally and $p_v$ as the probability of a rectangle being placed vertically, 
\begin{align}
p_h=n_{h}/N\\
p_v=n_{v}/N = 1 - p_h
\end{align}
Then, parameter $s$ can be written as,
\begin{align}
s= |p_v-p_h|
\end{align}

We can write the average excluded volume \cite{meeks} of our system of multi-oriented rectangles as, 
\begin{equation}
V_{ex}=p_vp_v V_{ex}^{vv}+p_hp_hV_{ex}^{hh}+p_vp_hV_{ex}^{vh}+p_hp_v V_{ex}^{hv}
\label{equation7}
\end{equation}

where $V_{ex}^{vv}$ is the excluded volume of a vertical rectangle to another vertical rectangle (See Sec.\ref{subsec1} for the definition of connectedness between shapes), $V_{ex}^{hh}$ is the excluded volume of a horizontal rectangle to another horizontal rectangle, and $V_{ex}^{hv} = V_{ex}^{vh} $ is the excluded volume of a vertical rectangle to a horizontal one. 

For multi-oriented rectangles of length $k_1$ and width $k_2$, we have
\begin{align*}
 V_{ex}^{vv}=(2k_1+1)(2k_2+1)-5\\
  V_{ex}^{hh}=(2k_1+1)(2k_2+1)-5,\\
  V_{ex}^{vh}=(k_1+k_2+1)^2-4,\\
  V_{ex}^{hv}=(k_1+k_2+1)^2-4,
\end{align*}
Using these in \ref{equation7}, we get 
\begin{multline}
V_{ex}=\left(\frac{1+s}{2}\right)^2((2k_1+1)(2k_2+1)-5)+\\ \left(\frac{1-s}{2}\right)^2((2k_1+1)(2k_2+1)-5)+\\ \frac{1-s^2}{4}((k_1+k_2+1)^2-4)+\frac{1-s^2}{4}((k_1+k_2+1)^2-4)
\end{multline}
Which can be used in the expression for $\phi_c$,
\begin{equation}
\phi_c=1-\exp\left(-B_c\frac{V}{V_{ex}(s)}\right)
\end{equation}
For $s=1$, we recover the results described in Sec.\ref{subsec1}. For all values of $s<1$, from the above expression, we can easily verify that $\phi_c$ decreases monotonically for large $k_1$. The ratio $\frac{V}{V_{ex}}$ will depend inversely on $k_1$ for large $k_1$ and $k_1 >> k_2$. Thus the peculiar behavior observed for $s=1$ (fully aligned) case is unique.


\begin{thebibliography}{}

\bibitem{sahimi}M. Sahimi. \textit{Applications of Percolation Theory}, Taylor and 
Francis, London (1994)
\bibitem{sander}L. M. Sander, C. P. Warren, and I. M. Sokolov, Epidemics, disorder, and percolation, Physica A 325, 1 (2003).
\bibitem{jiantong}Jiantong Li, Zhi-Bin Zhang and Shi-Li Zhang, Percolation in
random networks of heterogeneous nanotubes, Appl. Phys. Lett. 91:25, 253127 (2007).
\bibitem{li}C. Li and T.-W. Chou, Continuum percolation
of nanocomposites with fillers of arbitrary shapes, Appl. Phys. Lett. 90, 174108 (2007).
\bibitem{drossel}B. Drossel and F. Schwabl, Forest-fire model with immune
trees, Physica A 199, 183 (1993).
\bibitem{klypin}A. A. Klypin. and S. F. Shandarin., Percolation technique
for galaxy clustering, ApJ, 413, 48 (1993).
\bibitem{adam}M. Adam, M. Delsanti, J.P. Munch, and D, Durand, Sol-gel transition: A model for percolation, Physica A 163, 85 (1990).
\bibitem{stauffer1}D. Stauffer, Percolation models of financial market dynamics. Advances in complex systems,04(01):19-27 (2001).
\bibitem{christensen} K. Christensen. and N. R. Moloney, \textit{Complexity and Criticality}(Imperial College Press, 2005).
\bibitem{stauffer}  D. Stauffer and A. Aharony, \textit{Introduction to Percolation Theory} (Taylor and Francis, London, 2018).
\bibitem{mertens}  S. Mertens and C. Moore, Continuum percolation thresholds in two dimensions, Phys. Rev. E 86, 061109 (2012).
\bibitem{xia} W. Xia. and M. F. Thorpe., Percolation properties of random
Ellipses, Phys. Rev. A 38, 2650 - 6 (1988).
\bibitem{baker}D. R. Baker, G. Paul, S. Sreenivasan, and H. E. Stanley, Continuum percolation threshold for interpenetrating squares and cubes,  Phys. Rev. E, 66:046136 (2002).
\bibitem{gawlinski}E. T. Gawlinski and H. E. Stanley, Continuum percolation
in two dimensions: Monte Carlo tests of scaling and universality for non-interacting discs, J. Phys. A: Math. Gen. 14, L291 (1981).
\bibitem{becklehimer} J. Becklehimer and R. B. Pandey, A computer simulation study of sticks percolation, Physica A 187, 71 (1992).
\bibitem{vandewalle}N. Vandewalle, S. Galam, and M. Kramer, A new universality
for random sequential deposition of needles, Eur. Phys. J. B 14, 407 (2000).
\bibitem{longone} P. Longone, P.M. Centres, A.J. Ramirez-Pastor, Percolation
of aligned rigid rods on two-dimensional square lattices, Phys. Rev. E 85, 011108 (2012).
\bibitem{tarasevich}Y. Y. Tarasevich and A. V. Eserkepov, Percolation of sticks: Effect of stick alignment and length dispersity, Phys.Rev. E 98, 062142 (2018).

\bibitem{tarasevich1} Y. Y. Tarasevich, N. I. Lebovka, and V. V. Laptev, Percolation of linear k-mers on a square lattice: From isotropic through partially ordered to completely aligned states, Phys. Rev. E 86, 061116 (2012).

\bibitem{cornette} V. Cornette, A. J. Ramirez-Pastor, and F. Nieto, Dependence
of percolation threshold on the size of percolating species, Physica A 327, 71 (2003).

\bibitem{cornette1} V. Cornette, A. J. Ramirez-Pastor, F. Nieto, Percolation
of polyatomic species on a square lattice, Euro. Phys. J. B 36, 391-399 (2003).

\bibitem{lebrecht}W. Lebrecht, J. F. Vald´es, E. E. Vogel, F. Nieto, and A. J.
Ramirez-Pastor, Percolation of dimers on square lattices, Physica A 392, 149 (2013).

\bibitem{grzegorz} G. Kondrat, A. Pekalski, Percolation and
jamming in random sequential adsorption of linear segments on a square lattice, Phys. Rev. E 63, 051108 (2001).
\bibitem{grzegorz1} G. Kondrat, Impact of composition of extended objects on percolation on a lattice, Phys. Rev. E 78, 011101 (2008).
\bibitem{cherkasova}V. A. Cherkasova, Y.Y. Tarasevich, N. I. Lebovka, and N. V. Vygornitskii, Percolation of aligned dimers on a square lattice, Eur. Phys. J. B 74,205 (2010).
\bibitem{longone1} P. Longone, P. M. Centres, and A. J. Ramirez-Pastor, Percolation of aligned rigid rods on two-dimensional triangular lattices, Phys. Rev. E 100, 052104 (2019).
\bibitem{koza}Z. Koza, G. Kondrat, and K. Suszczy´nski, Percolation
of overlapping squares or cubes on a lattice, J. Stat. Mech. 
 Theory Exp., 2014(11):P11005, (2014)
\bibitem{koza1}Z. Koza, J. Pola, From discrete to continuous percolation
in dimensions 3 to 7, J. Stat. Mech. Theory Exp., 2016,103206 (2016).
\bibitem{brzeski}P. Brzeski, G. Kondrat, Percolation of hyperspheres in dimensions 3 to 5: from discrete to continuous, J. Stat. Mech. Theory Exp., 2022,053202 (2022).

\bibitem{mecke} K. R. Mecke and A. Seyfried, Strong dependence of percolation thresholds on polydispersity, EPL-Europhys. Lett.,58(1):28 (2002).

\bibitem{xun}Z. Xun, D. Hao, and R. M. Ziff, Site percolation on square
and simple cubic lattices with extended neighborhoods and their continuum limit, Phys. Rev. E 103, 022126 (2021).

\bibitem{majewski} M. Majewski and K. Malarz, Square lattice site percolation
thresholds for complex neighborhoods, Acta Phys. Pol. B 38 2191 (2007).


\bibitem{alvarez}A. Alvarez-Alvarez, I. Balberg, and J. P. Fernandez-Alvarez, Invariant percolation properties in some continuum systems, Phys. Rev. B, 104, 184205(2021).

\bibitem{balberg2}I. Balberg, The physical fundamentals of the electrical conductivity in
nanotube-based composites, J. Appl. Phys. 128, 204304 (2020).

\bibitem{du} F. Du, J. E. Fischer, and 
 K. I. Winey, Effect of nanotube
alignment on percolation conductivity in
carbon nanotube/polymer composites, Phys. Rev B 72,121404(R) (2005).

\bibitem{ni} X. Ni, C. Hui, N. Su, W. Jiang, and F. Liu, Monte Carlo simulations of electrical percolation in multicomponent thin films with nanofillers, Nanotechnology 29,075401 (2018).

\bibitem{ramasubramaniam}R. Ramasubramaniam, J. Chen, and H. Liu, Homogeneous carbon nanotube/polymer composites for electrical applications, Appl. Phys. Lett. 83, 2928 (2003).

\bibitem{hu} L. Hu, D. S. Hecht, and G. Gruner, Percolation in Transparent and Conducting Carbon Nanotube Networks, Nano Lett. 4, 2513(2004).

\bibitem{berhan}L. Berhan and A. M. Sastry, Modeling percolation in high-aspect-ratio fiber systems. I. Soft-core versus hard-core models,  Phys Rev E 75, 041120 (2007).

\bibitem{berhan1} L. Berhan and A. M. Sastry, Modelling percolation in high aspect-ratio fiber systems. II. The effect of waviness on the percolation onset, Phys. Rev. E 75, 041121(2007).




\bibitem{adler}P. M. Adler, Transport processes in fractals. Stokes flow through Sierpinski carpets, Phys. Fluids 29,15-22 (1986).

 \bibitem{saleh} S. Saleh, J. E Thovert and P.M. Adler, Measurement of two-dimensional velocity fields in porous media by particle image displacement velocimetry Experiments in Fluids 12 (1992).
 
\bibitem{yaofa} Yaofa Li, Farzan Kazemifar, Gianluca Blois, Kenneth T. Christensen, Micro-PIV measurements of multiphase flow of water and liquid CO2 in 2-D heterogeneous porous micromodels, Water Resources
Research, 53(7), 6178-6196 (2017).

 \bibitem{tang} L. P. Tang, S. Yang, D. Liu, et al, Two-dimensional porous coordination polymers and nano-composites for electrocatalysis and electrically conductive applications, Journal of Materials Chemistry A 8.29: 14356-14383(2020).

\bibitem{koponen} A. Koponen, M. Kataja, and J. Timonen, Tortuous flow in porous media, Phys. Rev. E 54, 406(1996).

\bibitem{koponen1} A. Koponen, M. Kataja, J. Timonen, Permeability and effective porosity of porous media, Phys. Rev. E 56, 3319-3325(1997).

\bibitem{koponen2} A. Koponen.,M. Kataja, J. Timonen, Simulations of single fluid flow in porous medium, Int. J. Mod. Phys. C 9, 1505-1521(1998).

\bibitem{matyka} M. Matyka, A. Khalili,Z. Koza,  Tortuosity-porosity relation in porous media flow. Phys. Rev. E 78(2), 026306(2008).

\bibitem{koza2} Z.  Koza,  M.  Matyka,  A.  Khalili,  Finite-size  anisotropy  in  statistically  uniform  porous  media. Phys.  Rev.  E  79,  066306(2009).

\bibitem{mecke1} Mecke K.R, Additivity, convexity, and beyond: Applications of Minkowsky functionals in statistical physics. In \textit{Statistical physics and spatial statistics}, Springer:Berlin/Heidelberg,Germany, New York,NY,USA;pp. 111-184 (2000).


\bibitem{klatt} M. A. Klatt, G. E. Schroder-Turk, and K. Mecke, Anisotropy in finite continuum percolation: Threshold estimation by Minkowsky functionals, J. Stat. Mech. Theory Exp. 2017(2), 023302 (2017).

\bibitem{dhar} D. Dhar, On the critical density for continuum percolation of spheres of variable radii, Physica A 242, 341-346 (1997).



\bibitem{balberg}I. Balberg and C. H.  Anderson, S. Alexander, and N. Wagner, Excluded volume and its relation to the onset of percolation, Phys. Rev. B 30, 3933-43 (1984).

\bibitem{balberg1} I. Balberg, Recent developments in continuum percolation, Phil. Mag B 56, 991-1003 (1987).

\bibitem{xun1} Z. Xun, D. Hao, and R. M. Ziff, Site and bond percolation on regular lattices with compact extended-range neighborhoods in two and three dimensions, Phys. Rev. E 105, 024105 (2022).

\bibitem{johnston} Johnston L. Bernard, Richman Fred, \textit{Numbers and Symmetry: An Introduction to Algebra} (1997).

\bibitem{gouker}M. Gouker and F. Family, Evidence for classical critical behavior in long-range site percolation, Phys. Rev. B 28, 1449 (1983).


\bibitem{hoshen}J. Hoshen and  R. Kopelman, Percolation and cluster distribution. I. Cluster multiple labeling technique and critical concentration algorithm, Phys. Rev. B 14,3438 (1976).

\bibitem{rintoul} M. D. Rintoul and S. Torquato, Precise determination of the critical threshold and exponents in a three dimensional continuum percolation model, J. Phys. A:  Math. Gen. 30, L585, (1997).

\bibitem{fabian} F. Coupette, R. de Bruijn, P. Bult, S. Finner, M. A. Miller, P. van der Schoot, and  T. Schilling, Nearest-neighbor connectedness theory: A general approach to continuum percolation, Phys. Rev. E 103, 042115(2021).

\bibitem{fabian1} F. Coupette and  T. Schilling, Exactly solvable percolation problems, Phys. Rev. E 105, 044108 (2022).

\bibitem{fabian2} F.Coupette and Tanja Schilling, Universal approach to critical percolation, arXiv:2308.16757.

\bibitem{mecke2}Mecke K.R, Integral geometry in statistical physics, Int. J. Mod. Phys. B,12, 861-899 (1998).

\bibitem{sasidevan}  V. Sasidevan, Continuum percolation of overlapping disks with a distribution of radii having a power-law tail, Phys. Rev. E 88, 022140 (2013).
\bibitem{hall} P. Hall, On continuum percolation, Ann. Probab. 13(4), 1250-1266 (1985).

\bibitem{bai}T. Bai, D. D. Pollard, H. Gao, Explanation for fracture spacing in layered materials, Nature 403,753-756(2000).

\bibitem{xu} D. Xu, T. Han and L.-Y. Fu, Seismic dispersion and attenuation in layered porous rocks with fractures of varying orientations, Geophys. Prospect., 69, 220-235(2021).

\bibitem{meeks} K. Meeks, M. L. Pantoya, M. Green, and J. Berg, Extending the excluded volume for
percolation threshold estimates in polydisperse systems: The binary disk system, Appl.
Math. Model., vol. 46, pp. 116-125,(2017).


\end{thebibliography}
\end{document}